# MAGNETO: Covert Channel between Air-Gapped Systems and Nearby Smartphones via CPU-Generated Magnetic Fields

Mordechai Guri, Andrey Daidakulov, Yuval Elovici
Ben-Gurion University of the Negev
Cyber Security Research Center
(gurim@post.bgu.ac.il; daydakul@post.bgu.ac.il; elovici@post.bgu.ac.il)
**Video: https://cyber.bgu.ac.il/advanced-cyber/airgap**

*Abstract—* In this paper, we show that attackers can leak data from isolated, air-gapped computers to nearby smartphones via covert *magnetic* signals. The proposed covert channel works even if a smartphone is kept inside a Faraday shielding case, which aims to block any type of inbound and outbound wireless communication (Wi-Fi, cellular, Bluetooth, etc.). The channel also works if the smartphone is set in 'airplane mode' in order to block any communication with the device.

We implement a malware that controls the magnetic fields emanating from the computer by regulating workloads on the CPU cores. Sensitive data such as encryption keys, passwords, or keylogging data is encoded and transmitted over the magnetic signals. A smartphone located near the computer receives the covert signals with its magnetic sensor. We present technical background, and discuss signal generation, data encoding, and signal reception. We show that the proposed covert channel works from a user-level process, without requiring special privileges, and can successfully operate from within an isolated virtual machine (VM).

I. INTRODUCTION

Magnetic sensors (also known as magnetometers) in smartphones and tablets are mainly used for orientation and positioning. Unlike cellular, Wi-Fi, Bluetooth, and NFC hardware, magnetic sensors are not considered communication interfaces. Thus, they can be accessed with basic permissions and remain active even in airplane mode, where all wireless communication interfaces are disabled.

In this paper, we show how attackers can maliciously utilize a smartphone's magnetic sensors in order to exfiltrate information from isolated, non-networked (air-gapped) computers. In this attack, a malware on the air-gapped computer gathers sensitive data and transmits it via magnetic fields generated from the CPU cores. The covert magnetic signals are then received by a smartphone located in close proximity to the computer. The data received is decoded and sent to the attacker via the Internet (e.g., Wi-Fi or cellular data).

### A. Air-Gap Covert Channels

An air-gapped network is a secured computer network in which security measures are taken to maintain physical separation from public networks such as the Internet. Such security measures taken may include separation of network equipment, using physical access control to systems, banning unauthorized hardware and so on. But despite of these restrictions, several incidents in the last decade has shown that air-gapped networks are not immune to malware infections [1][2][3]. For example, In 2017, WikiLeaks published a reference to a tool dubbed 'Brutal Kangaroo,' allegedly used to infiltrate air-gapped computers via external USB drives [4]. Using such techniques, attackers can breach the network and bypass security measures such intrusion detection and intrusion prevention systems (IDS/IPS). In order to *exfiltrate* data from the compromised air-gapped, networkless, computer, attackers must leak the data through special types of covert channels known as *air-gap covert channels*. Over the years, several types of air-gap covert channels have been proposed based on electromagnetic [5–9], acoustic [10] [11] [12], optical [13] [14], and thermal [15] emanations from the computer. In this paper, we propose a magnetic covert channel between air-gapped computers (e.g., workstation) and smartphone located nearby. Note that the covert channel requires that malicious code is run on both sides: on the air-gapped computer and the adjacent smartphone.

### B. Our Contribution

We introduce a new type of covert communication between air-gapped computers and nearby smartphones. Although the proposed covert channel is limited in terms of distance and speed, it has unique characteristics:

**Availability.** The hardware used for the covert communication exists in virtually all modern computers and smartphones; the CPU used for the transmission is a fundamental component of any computer, and the magnetic sensors used for the reception are an integral part of any modern smartphone or tablet.

**Required privileges.** The malicious codes executed in the computer and smartphone require no special privileges (e.g., root or admin) and can operate from an ordinary user-level process or app.

The proposed covert channel even functions in constrained environments where other types of wireless communication are blocked:

- **Faraday shielding.** Computers and smartphones in secure areas might be kept in Faraday shielding or a Faraday case to prevent any type of wireless communication to/from the device. Nevertheless, the low frequency magnetic fields used in our covert channel can bypass Faraday shielding.

- **Airplane mode.** In some secure zones a smartphone might be set in airplane mode. In this mode all types of wireless communication interfaces in the device (Wi-Fi, cellular, Bluetooth, etc.) are disabled, however the magnetic sensors in smartphones are not considered communication interfaces, and hence remain active even in airplane mode.

The rest of the paper is structured as follows: In Section II we present related work. Section III provides technical background. Section IV describes the attack model. The transmission and reception are described in Sections V and VI. In Section VII we present the analysis and evaluation. We discuss countermeasures in Section VIII, and we conclude in Section IX.

## II. RELATED WORK

Air-gap covert channels can be classified into five main categories: electromagnetic, magnetic, acoustic, thermal, and optical.

In the past twenty years, several studies have proposed the use of electromagnetic emanation from computers for covert communication. In 1998, Kuhn and Anderson [6] showed that attackers can control the electromagnetic emissions from computer displays. In this study a malicious code generated radio signals in the AM band and encoded binary data on them. Thiele [16] presented a program dubbed 'Tempest for Eliza,' which used the VGA cable to transmit the song 'letter to Alice' played on the AM radio signals. In 2014, Guri et al introduced AirHopper [5], [17], malware that can leak data from air-gapped networks to a nearby smartphone using electromagnetic signals in the FM radio band emanating from the video cable. Guri et al also presented GSMem [9], malware that leaks data from air-gapped computers using frequencies in the GSM, UMTS, and LTE band emitted from the RAM buses. Their method used a multi-channel memory architecture to generate an amplified signal, which was received by a rootkit placed on the baseband firmware of a compromised mobile phone. In 2016, Guri et al demonstrated a method dubbed USBee that uses the USB connectors to generate covert electromagnetic signals from PCs [18].

Sound waves offer another way of leaking data from air-gapped computers. Hanspach proposed using ultrasonic sound to transmit data between laptops with speakers and microphones [12]. However, this method is only relevant when speakers and microphones are present. In 2016, Guri et al demonstrated Fansmitter [10] and DiskFiltration [11], two methods enabling exfiltration of data via specially generated sound waves when the air-gapped computers are not equipped with speakers. The binary data is encoded on top of noise emitted from the PC fans and hard disk drive mechanical arm.

Data can also be leaked from air-gapped computers optically. Loughry and Umphress proposed using the PC keyboard LEDs (caps-lock, num-lock and scroll-lock) to modulate data [19]. In 2017, Guri et al presented a covert channel that uses the hard drive indicator LED [13] and router LEDs [20] in order to leak data from air-gapped computers and networks. VisiSploit [21] is another optical covert channel in which data is leaked through fast blinking images or low contrast bitmaps projected on the computer screen. Lopes et al proposed using external USB devices implanted with IR LEDs to exfiltrate data [22]. In 2017, Guri et al presented aIR-Jumper, a method that uses the IR LEDs in security cameras to exfiltrate and infiltrate air-gapped networks remotely [14]. Their method enabled bidirectional communication between a remote attacker and malware inside the organizational network.

In 2015, Guri et al demonstrated a thermal covert channel referred to as BitWhisper [15]. In this technique, an attacker can establish bidirectional communication between two adjacent air-gapped computers using temperature manipulation. The heat is generated by the CPU of one PC and received by temperature sensors that exist in the motherboard of the other PC.

### A. Magnetic

Myhayun et al suggested using magnetic hard disk drives to generate magnetic emission, which can be received by a nearby smartphone magnetic sensor [23]. The smartphone needs to be located just a few centimeters from the transmitting laptop, and the bitrate varies from 0.067 bit/sec to 2 bit/sec. Recently, Guri et al presented ODINI [24], a malware that can exfiltrate data from air-gapped computers via low frequency magnetic signals generated by the computer's CPU cores. They also showed that the low frequency magnetic fields can bypass Faraday cages. Their attack model requires an implanted magnetic receiver ('bug') in the proximity of the transmitting Faraday-caged air-gapped computer. The attack model presented in our paper is similar to [23] in terms of the reception, since it exploits a nearby infected smartphone. However, in the current research we adapt the signal generation process used by Guri et al in

[24]. The specific ways in which our method differs from [23] and [24] are listed in Table 1.

**Table 1. Comparison with other magnetic covert channels**

| Work | Signal generation | Receiver | Max bit-rate | Max distance |
|---|---|---|---|---|
| ODINI [24] | CPU cores | magnetic sensor | 40 bit/sec | 100 to 150 cm |
| MAGNETO (this paper) | CPU cores | smartphone | 5 - 0.2 bit/sec | 0 to 12 cm (desktops) |
| Hard Disk Drive ([23]) | Hard Disk Drive (magnetic) | smartphone | 2 - 0.06 bit/sec | 0 to 12 cm (laptops) |

Unlike [23] which requires a magnetic hard disk drive, our method use the CPU hence works on any computer. In addition, in this paper we focus on an attack model relevant for air-gapped workstations. Thus, we evaluated the covert channel on desktop workstations rather than laptop computers used in [23].

Table 2 summarizes the existing air-gap covert channels, including the magnetic ones.

**Table 2. Summary of existing air-gap covert channels**

| Type | Method |
|---|---|
| Electromagnetic | AirHopper [4], [11] <br> GSMem [15] <br> USBee [24] <br> Funthenna [25] |
| Magnetic | MAGNETO (this paper) <br> ODINI [24] <br> Myhayun (hard disk drive [38]) |
| Acoustic | Fansmitter (computer fan noise) [27] <br> DiskFiltration (hard disk noise) [28] |
| Thermal | BitWhisper [35] |
| Optical | LED-it-GO (hard drive LED) [30] <br> VisiSploit (invisible pixels) [32] <br> Keyboard LEDs [29] <br> Router LEDs [31] |
| Infrared (IR) | aIR-Jumper (security cameras & infrared) [34] <br> Implanted infrared LEDs [33] |

III. TECHNICAL BACKGROUND

Magnetic fields are produced whenever a charge is in motion, e.g., when current flows through a wire. A magnetic field at a given point is specified by its direction and strength, and is mathematically represented by a vector field. The intensity of magnetic fields is measured in teslas (T). One tesla is defined as the field intensity generating one newton (N) of force per ampere (A) of current per meter of conductor. Note that a magnetic field of one tesla is very strong, and magnetic fields are commonly measured in units of milliteslas ($1mT = 10^{-3}T$) or microteslas ($1\mu T = 10^{-9}T$).

Ampère's Law shows that the strength of a magnetic field is proportional to the current flow in a wire. The main disadvantage of the magnetic field is that the strength of a magnetic field is inversely proportional to the third power of the distance ($1/r^3$) from the magnet's center [25]. This characteristic limits the distance of magnetic communication compared to the distance achieved with electromagnetic communication [26]. In practice, magnetic fields are used for the establishment of short-range wireless communication between devices that are close to one another - a technique commonly referred to as Near-Field Magnetic Induction communication [27].

A. *Smartphone Magnetomers*

Magnetic sensors (magnetometers) are integrated into almost all modern smartphones and tablets. Their main functionality is to measure the earth's magnetic fields in order to detect the orientation of the phone in a three-dimensional space. A typical magnetometer contains three magnetic sensors (for the three perpendicular axes X, Y, and Z). Magnetometer chips commonly include a built-in accelerometer that helps adjust the measurements of the magnetic sensor.

There are a few types of magnetometers used in consumer-grade smartphones. The Hall effect sensor is the most commonly used magnetometer in smartphones and tablets [28]. This kind of sensor produces voltage when the sensor is placed in a strong magnetic field perpendicular to its plane. The voltage produced is proportional to the strength of the magnetic field. The voltage sensed is then converted to a signal representing the magnetic field intensity in telsas. Other types of magnetometers that are less common in mobile devices include the giant magnetoresistance (GMR), anisotropic magnetoresistance (AMR), and the magnetic tunneling junction (MTJ) magnetometers. The work in [29] provides a comprehensive explanation of the principles of these magnetic sensors. The magnetometers in modern smartphones have a resolution in the tens of μT and sampling rates between 50Hz and 150Hz.

IV. ATTACK MODEL

The attack model consists of three steps: infection of an air-gapped network, infection of a smartphone, and data exfiltration.

In the first step, one node in the air-gapped network is contaminated, possibly through removable media, outsourced software, or hardware components. Note that several of the APTs discovered in the last decade were capable of infecting air-gapped networks [30] (e.g., Turla [31], Red October [32], and Fanny [33]). The malware should then propagate itself across the network, possibly targeting specific computers according to a given criteria. In the second step, the smartphone of a targeted employee is located, and the receiver application is installed on the smartphone. The device can be infected via email attachments, compromised websites, or malicious app downloads. When the first two steps are accomplished, the attack enters its exfiltration phase. Although employees are assumed to carry their mobile phones around the workplace, the devices are often placed on a desk near the user's computer. The malicious program lies undetected within the targeted computer, gathering sensitive data when possible. The data might be keylogging data, encryption keys, credential tokens, or passwords. The malware then leaks the data via the magnetic emanation generated from the CPU. Concurrently, the malicious app on the phone scans the magnetic field for the signal denoting an incoming data transmission. When the signals are received and decoded, the data can be forwarded by the smartphone to the attacker's computer encrypted via Wi-Fi.

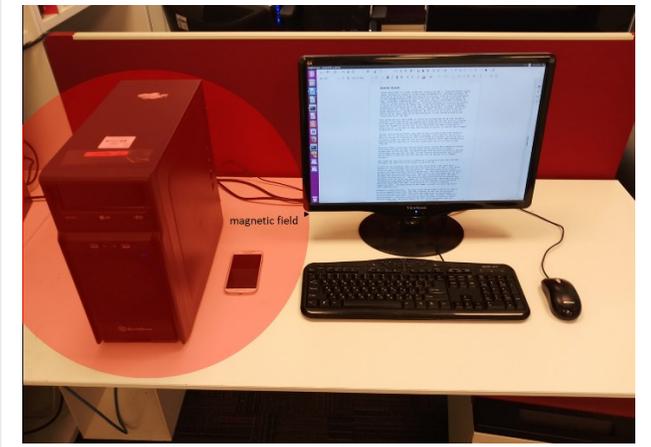

**Figure 1. An exfiltration scenario in which data is modulated over a covert magnetic field emanating from the air-gapped computer; the data is received by a nearby smartphone.**

It is important to note that the mobile phone's owner is not necessarily a malicious insider; the device may be contaminated by an "evil maid" attack in which an individual tampers with a device temporarily left unattended, without the owner's consent or knowledge.

Figure 1 presents a typical leakage scenario in which an employee's smartphone is placed on the desk near the infected computer. Covert magnetic signals are transmitted over the air-gap and received by the malicious app in the smartphone. Data is decoded and sent encrypted to the attacker via Wi-Fi.

## V. Transmission

As described in the technical background section, moving charges in a wire generate a magnetic field. The magnetic field changes according to the acceleration of the charges in the wire. In a standard computer, the wires that supply electricity from the main power supply to the motherboard are the primary source of the magnetic emanation. The CPU is one of the largest consumers of power in the motherboard. Since modern CPUs are energy efficient, the momentary workload of the CPU directly affects the dynamic changes in its power consumption [34]. By regulating the workload of the CPU, it is possible to govern its power consumption, and hence to control the magnetic field generated. In the most basic case, overloading the CPU with calculations will consume more current and generate a stronger magnetic field. By intentionally starting and stopping the CPU workload, we can generate a magnetic field at the required frequency and modulate binary data over it.

To generate a carrier wave at frequency $f_c$ in one or more cores, we control the utilization of the CPU at a frequency correlated to $f_c$. To that end, $n$ worker threads are created, where each thread is bound to a specific core. To generate the carrier wave, each worker thread repeatedly overloads its core at frequency $f_c$. That is, each thread applies a continuous workload on its core (full power consumption) for a time period of $1/2f_c$ and then puts its core in an idle state (low power consumption) for a time period of $1/2f_c$. The basic operation of a worker thread is described in Algorithm 1.

```
Algorithm 1
 1: procedure WORKERTHREAD(iCore, freq, nCycles0, nCycles1)
 2:     bindThreadToCore(iCore)
 3:     half_cycle_ms ← 0.5 * 1000/freq
 4:     while (!endTransmission()) do
 5:         if (data[i] = 0) then
 6:             sleep(nCycles0*half_cycle_ms*2)
 7:         else
 8:             for j ← 0 to nCycles1 do
 9:                 T1 ← getCurrentTime()
10:                 while (getCurrentTime() − T1 < half_cycle_ms) do
11:                     ;
12:                 end while
13:                 sleep(half_cycle_ms)
14:             end for
15:         end if
16:     end while
17: end procedure
```

A worker thread receives the core to be bound to (*iCore*) and the carrier frequency (*freq*). It also receives the number of cycles for the modulation of the logical '0' (*nCycles0*) and logical '1' (*nCycles1*). Note that the cycle time is derived from the frequency of the carrier wave (line 3). The thread function iterates on the array of bits to transmit. In the case of a logical '0' it sleeps for *nCycles0* cycles. In the case of a logical '1' it repeatedly starts and stops the workload of the core at the carrier frequency *freq* for *nCycles1* cycles. We overload the core using the busy waiting technique. This function causes full utilization of the core for the *time period* using a busy loop and then returns.

### A. Data Modulation

We implemented two data modulation schemes for the transmission which are described below: on-off keying (OOK) and binary frequency-shift keying (B-FSK). We denote the number of cores available for the transmission as $N_c$.

#### 1) On-off keying

In OOK modulation, the data is represented by the presence and absence of the carrier wave. The presence of a carrier wave represents the symbol '1,' while its absence represents the symbol '0' (Table 3). Note that in our covert channel, the amplitude of the carrier wave is unknown to the receiver in advance, and it mainly depends on the type of transmitting computer used, the number of cores participating in the transmission, and the distance between the transmitter and the receiver. These parameters are synchronized with the receiver during the preamble and bit-framing described later.

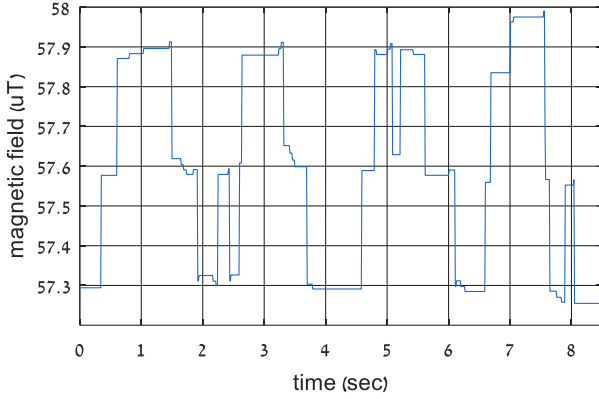

**Figure 3. The waveform of a binary sequence ('10101010') modulated with OOK.**

**Table 3. On-off keying**

| Symbol | Carrier wave | # of cores |
|---|---|---|
| 0 | Present | $n \leq N_c$ |
| 1 | Absent | $n \leq N_c$ |

Error! Reference source not found. shows the waveform of a binary sequence ('10101010') modulated with OOK and transmitted from a desktop PC with four cores. In this case, the smartphone was located at a distance of 5cm from the transmitting computer; the noise level is ~57.3µT, and the carrier wave amplitude is ~57.9µT.

*1) Binary Frequency-Shift Keying*

In frequency-shift keying (FSK) the data is represented by a change in the frequency of a carrier wave. Recall that the transmitting code can determine the frequency of the signal by setting the cycle time in the signal generation algorithm. In FSK, each frequency is represented by a different symbol. Table 4 contains a case in which we encode the two symbols '0' and '1' by the two frequencies $F_0$ and $F_1$.

**Table 4. Frequency-shift keying**

| Symbol | Frequency | # of cores |
|---|---|---|
| 0 | $F_0$ | $n \leq N_c$ |
| 1 | $F_1$ | $n \leq N_c$ |

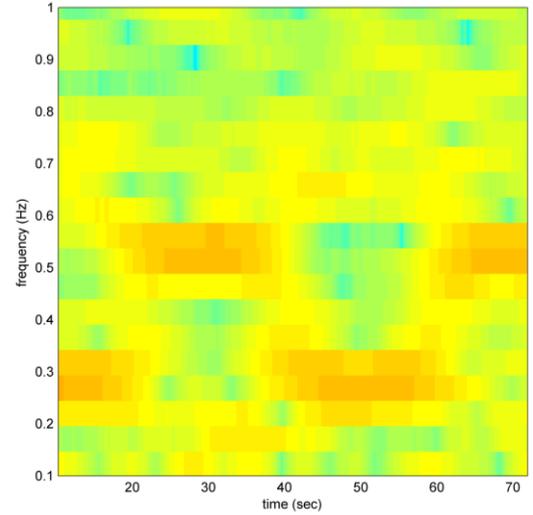

**Figure 2. FSK modulation of a binary sequence ('1010') with two frequencies (0.25Hz and 0.5Hz).**

Figure 2 shows the time-frequency spectrogram of a binary sequence ('0101') modulated with FSK as transmitted from a PC with four cores. In this modulation, the frequencies 0.25Hz and 0.5Hz have been used to encode the symbols '0' and '1' respectively. The receiving smartphone is located 5cm from the transmitting computer.

*Bit-Framing*

The data is delivered in small frames of 40 bits comprised of a preamble, a payload, and a CRC, as shown in Figure 4.

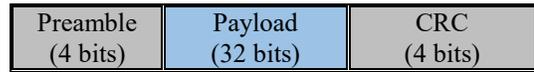

**Figure 4. Bit-framing.**

- **Preamble**. The preamble is transmitted at the beginning of every packet. It consists of a sequence of four alternating symbols ('1010') which helps the receiver determine the properties of the channel, such as the carrier wave frequency and amplitudes. Note that these properties are dependent on the distance of the smartphone from the transmitter, and hence might change during the transmission. In addition, the preamble header allows the receiver to detect the beginning of the transmission of each packet.

- **Payload**. The payload is the raw binary data to be transmitted. We choose 32 bits as the payload size.

- **CRC.** For error detection, we insert four bits of CRC code at the end of the frame. The receiver calculates the CRC for the received payload, and if it differs from the received CRC, an error is detected.

## VI. RECEPTION

In this section, we briefly discuss the implementation details and design considerations for the receiver side.

We implemented the receiver as an Android application. We access the magnetic sensor through the `android.hardware.Sensor` class [35] to create an instance of the `Sensor.TYPE_MAGNETIC_FIELD`. The application samples the magnetic field with the fastest sampling rate available on the device (using the `SensorManager.SENSOR_DELAY_FASTEST` settings). The magnetic sensor provides the field strength in three axes $(x, y, z)$; we calculate the field strength by the norm of the vector $\sqrt{(x^2 + y^2 + z^2)}$. Reception of the transmitted data is based on three main steps:

- **Signal Sampling**. Sampling the magnetic field and filtering the signal.
- **Preamble Detection.** Searching for a frame header.
- **Demodulation**. Demodulating the frame's payload.

The `SensorSample()` event handler is invoked when the sensor has new data. The handler has three possible states: SAMPLE, PREAMBLE, and DEMODULATE. In the SAMPLE state the code measures the magnetic field. In the PREAMBLE state it searches for a bit-frame header to identify the beginning of a new frame. In the DEMODUALTE state the actual signal demodulation occurs. The outline of the receiver app is presented in Algorithm 2.

```
Algorithm 2
 1: procedure SENSORSAMPLE(x, y, z)
 2:     raw_signal.add(√(x² + y² + z²))
 3:     signal ← FFT(raw_signal)
 4:     filtered_signal ← Filter(signal, Fc)
 5:     UpdateMovingAverage(filtered_signal)
 6:     if (state == PREAMBLE) then
 7:         if (DetectPreamble(filtered_signal) == true) then
 8:             SetState(DEMODULATE)
 9:         end if
10:     end if
11:     if (state == DEMODULATE) then
12:         bit ← Demodulate(filtered_signal)
13:         bitArray.add(bit)
14:         if ((bitArray.size%32==0) or SignalLost(filtered_signal)) then
15:             SetState(PREAMBLE)
16:         end if
17:     end if
18: end procedure
```

**Signal Sampling**. The first step is to measure the signal in a carrier wave $f_c$. Note that this step includes performing a fast Fourier transform on the sampled signal. Although we assume that the $f_c$ is known to the receiver in advance, more complicated implementation may include a preliminary phase for scanning the carrier wave $f_c$. The signal measured is stored in a buffer after applying a moving average filter for noise mitigation (lines 1-5). This data is used later in the demodulation routines. The noise mitigation function is applied to the current sample by averaging it with the last $W$ original samples.

**Preamble Detection.** In the PREAMBLE state the receiver searches for a preamble sequence to identify a frame header (lines 7-11). If the preamble sequence '1010' is detected, the state is changed to DEMODULATE to initiate the demodulation process (lines 12-18). Based on the preamble sequence, the receiver determines the '1' and '0' amplitude levels and the signals' duration.

**Demodulation**. In the DEMODULATION state the payload is demodulated given the signal parameters retrieved in the PREAMBLE state. The demodulated bit is added to the current payload. If the full payload is received, the algorithm moves back to the preamble detection state (PREAMBLE).

The function *SignalLost()* returns true if, during the data reception, the signal power measured is weaker than the amplitude of the '0's from the preamble for three seconds straight. In this case, any partially received data is discarded or marked appropriately, and the algorithm returns to the PREAMBLE state. Signal loss may occur if the malware stops the transmission (e.g., for stealth or due to the computer shutting down). Signal loss may also occur if the smartphone has been moved away from the computer and lost the signal.

## VII. ANALYSIS AND EVALUATION

In this section we present the evaluation of the covert channel. For testing we use the Samsung Galaxy S4 smartphone (GT-I9505). This model is shipped with the Yamaha YAS532B sensor [36] which is a 3-axis geomagnetic sensor chip integrated into the smartphone board. It has a sensitivity of 0.15 µT in the X and Y axes and a sensitivity of 0.25 µT in the Z axis. The experimental setup consists of a standard desktop PC as described in Table 5. The systems in the experiments were running Linux Ubuntu version 16.04 64-bit.

Table 5. Workstation used in the experiments

| # | Model | Motherboard/board | CPU | Cores |
|---|---|---|---|---|
| PC | Infinity desktop PC | Gigabyte H87M-D3H | Intel Core i7-4770 CPU @ 3.4GHz | 4 cores (8 virtual cores) |

### A. *Signal-to-Noise Ratio (SNR)*

In digital communication the signal-to-noise ratio (SNR) measures the signal strength relative to the background noise. The ratio is measured in decibels (dB) using a signal-to-noise ratio formula of SNR=$10 log_{10}(signal/noise)$. We measured the level of the signal and noise from a smartphone located near a transmitting computer (PC). In these tests, we transmitted a signal consisting of a sequence of alternating bits (010101…) for a duration of 20 seconds. We measured the signal from distances of 0 to 15cm.

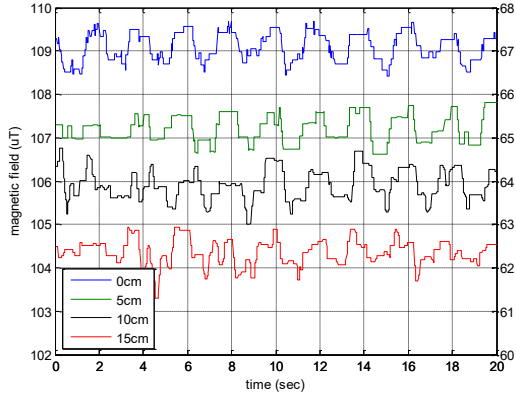

**Figure 5. Signal measurements at distances of 0 to 15cm.**

The measurement results are presented in Figure 5. As can be seen, the strength of the magnetic signal significantly decreases as the distance increases. At a distance of 0cm (the left Y-axis), the smartphone receives signals at levels of ~109.7μT (on average), while the background noise levels are ~108.5μT. At this distance the SNR level is 14.902dB. For distances of 5cm, 10cm, and 15cm (the right Y-axis), the SNR levels are 17.879dB, 12.420dB, and 10.240dB, respectively. For more than 15cm, the levels of the background noise were stronger than those of the signal; therefore, we stopped taking measurements at a distance of 15cm. Note that that SNR measured at close proximity (0cm) is lower than the SNR measured from 5cm. The lower SNR at 0cm is due to other magnetic fields emanating from the computer (e.g., the power supply), which interfere with our signal.

### B. *Bit Error Rate (BER)*

We measured the bit error rate (BER) of the covert channel for distances ranging from 0 to 15cm between a transmitting computer (PC) and a smartphone. For the BER measurements, we used simple OOK modulation. Our tests include three transmission rates: 0.2 bit/sec, 1 bit/sec, and 5 bit/sec. The results are shown in Table 6.

**Table 6. BER at various distances**

|  | 1cm | 5cm | 10cm | 12.5cm | 15cm |
|---|---|---|---|---|---|
| **0.2 bit/sec** | 0% | 0% | 0% | 10% | 30% |
| **1 bit/sec** | 0% | 0% | 20% | 20% | >30% |
| **5 bit/sec** | 0% | 20% | 30% | >30% | >30% |

We were only able to achieve a transmission rate of 5 bit/sec with 0% BER when the smartphone was in very close proximity of the transmitting computer (0-3 cm). Increasing the distance was found to compromise both the transmission rate and the BER; for example, at a distance of 12.5cm from the transmitting computer, the transmission rate of 1 bit/sec was accompanied by a high BER of 20%. With a slow transmission rate of 0.2 bit/sec, we were able to reach a distance of 12.5 cm with a BER of 10%. Based on these results we conclude that the covert channel is only relevant for the exfiltration of a small amount of data such as encryption keys and passwords. Given the limited distances and bitrates, a transmission rate of 1 bit/sec is the most practical among the various options given the trade-offs.

### C. *Interference*

The thread that generates the magnetic signals shares the CPU time with other processes in the operating system. We examined whether the activity of various processes interferes with signal generation. For this evaluation, we ran the transmitting process while employing four types of workloads common to desktop PCs which are described below:

- **Idle.** Only the default processes are running in the background.
- **Word processing.** The LibreOffice editor [37] is open, and the user is editing a document.
- **Video playing.** The VLC media player [38] is playing a movie clip.
- **CPU intensive calculations.** Linux matho-primes [39] is performing the calculations of large prime numbers.

**Table 7. SNR with various workloads**

| Workload | Application/Process | SNR (dB) |
|---|---|---|
| **Idle** | Background processes | 9.042 |
| **Word processing** | LibreOffice Writer | 8.990 |
| **Video playing** | VLC media player | 7.590 |
| **Calculations** | matho-primes | 3.430 |

Table 7 summarizes the SNR measured at a distance of 5cm from the transmitting computer for each of the four workloads. We used eight threads for the transmission of an alternating bit sequence ('10101010') using OOK modulation. As can be seen, the word processing and video playing result in just slightly more interference than that of the idle state. The main reason for this is the low amount of CPU utilization consumed by word processing and video playing. However, the intensive computational process associated with matho-primes results in a noticeable amount of background noise and reduces the SNR levels to ~3.4dB. Based on this we conclude that the covert channel will be more effective at times in which the transmitting computer is not performing CPU bound workloads.

### D. *Virtual Machines*

Virtualization technologies are widely used in modern IT environments (e.g., private and public clouds, and the virtualization systems of desktops and servers). One of the advantages of virtualization is the resource isolation it provides. Virtual machine monitors (VMM), also referred to as hypervisors, provide a separation between the guest operating system and the hardware resources. We examined the operability of a transmitter running in a virtualized environment with the primary goal of determining whether the virtualization layer causes interruptions or delays that affect signal generation.

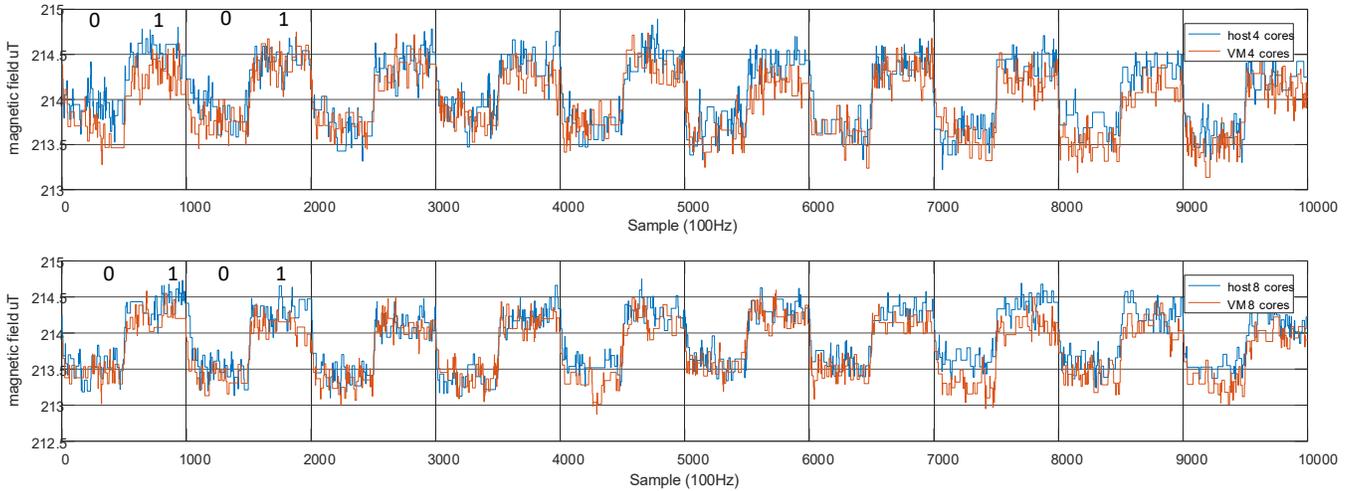

**Figure 6. The waveforms of the signals transmitted from a PC (VM / Host) with four cores (top) and eight cores (bottom)**

Figure 6-top (tests with four cores) and 6-bottom (tests with eight cores) show the waveforms of signals as measured by a smartphone located 3cm away from the PC. The blue signals were generated from the host computer, and the red signals were generated from within a VMware virtual machine. Both signals represent the transmission of the alternating sequence (101010…) using OOK modulation. The guest and host computers were running Linux Ubuntu 16.04 64-bit. We used VMware Workstation Player 14.0 for the virtualization and configured the host machine to support all available processors. As can be seen, the magnetic signals generated from within the VMs are similar to the magnetic signals generated directly from the host computer. For four cores, the SNR for both signals (host and VM) is about 6.3dB. For eight cores, the SNR for both signals is about 9dB. Note that we experienced no time delay or reduction in signal strength when transmitting from a VM.

### E. *Number of Cores*

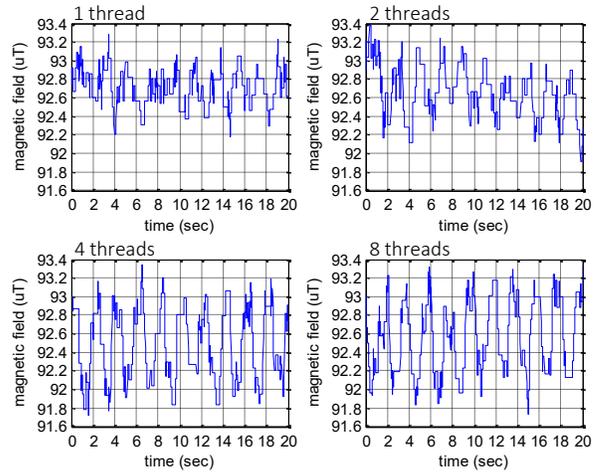

**Figure 7. The signal generated from 1, 2, 4, and 8 cores.**

As described in [24] the number of cores used for the transmission influences the currency flow, and hence the strength of the generated signal (i.e., the use of more transmitting cores yields a stronger signal). Figure 7 presents the waveforms of the alternating sequence (101010…) transmitted with 1, 2, 4, and 8 threads. In these tests, each thread was bound to a different core. As expected, the use of more cores yields higher SNR values: the measured SNR for 1, 2, 4, and 8 cores was 7.83dB, 10.64dB, 15.17dB, and 15.19dB, respectively.

### F. *Faraday Bags for Smartphones*

Faraday bags serve as small Faraday cages for handheld devices such as smartphones and tablets [40]. They are primarily aimed at blocking any inbound or outbound RF (radio frequency) activity to/from the device, including transmission and reception of cellular, Bluetooth, Wi-Fi, GPS, RFID, and NFC signals (Error! Reference source not found.). However, the magnetic covert channel proposed in this paper works even if the smartphone is kept inside a Faraday shielding case. In the proposed covert channel, we generate magnetic fields at frequencies lower than 50Hz. Such low frequency magnetic waves can propagate through metal, concrete, and soil [45][41]. As shown in [24] even the use of 3mm metal shields is ineffective for low frequency magnetic fields.

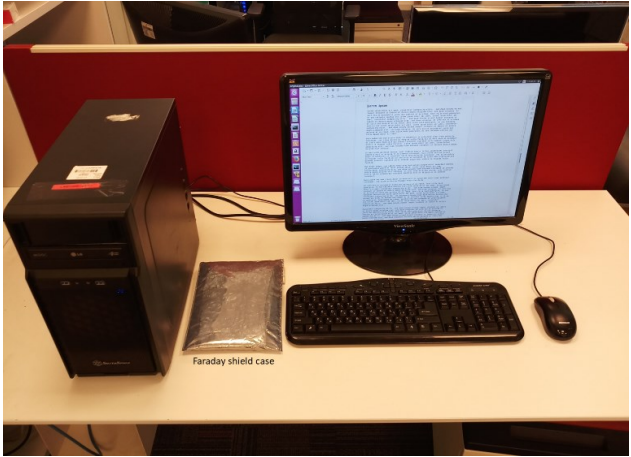

**Figure 9. A smartphone in a Faraday bag receives the magnetic signals from a transmitting computer nearby.**

Figure 8 shows the successful reception of a signal when the smartphone is stored inside and outside an enclosed faraday bag. In this test, the smartphone is placed 7 cm from the transmitting PC. As can be seen, the magnetic signals bypass the Faraday bag and are successfully received by the smartphone. The SNR is 5.6dB when the smartphone is placed inside the Faraday bag and 7.2dB when the smartphone is located outside the case.
`

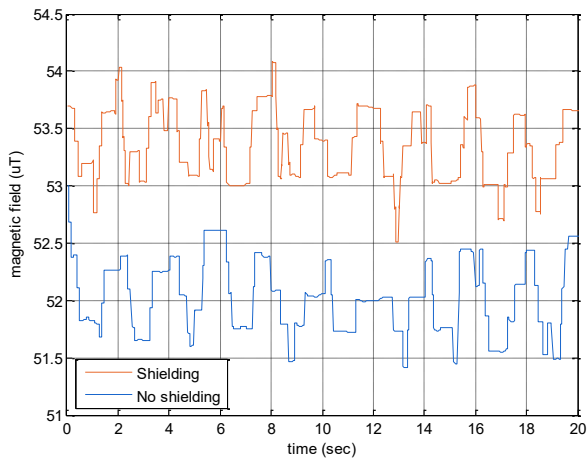

**Figure 8. The signal received with a smartphone placed inside and outside a Faraday bag**

### G. *Stealth*

The transmitting program leaves only a small footprint in the memory, making its presence easier to hide from AVs. At the OS level, the transmitting program requires no special or elevated privileges (e.g., root or admin), and hence can be initiated from an ordinary user space process. The transmitting code consists mainly of basic CPU operations such as busy loops, which do not expose malicious behavior, making it highly evasive from automated analysis tools. The receiver application on the smartphone side does not require root access and can be installed as a simple app with a basic permission to access the magnetic field sensor; this is true in both Google Android and Apple iOS devices [35][42]. With regard to the attack model, this means that the malicious app can be delivered via the app market. Such an app will likely be able to evade static and dynamic detection tools, since the access to the magnetic field sensor is considered safe in terms of privacy and security.

## VIII. COUNTERMEASURES

### A. *Detection*

Covert channel detection can be handled by security systems running in the computer or smartphone. In this approach, security solutions continuously trace the processes' activities and try to detect malicious operations. In the case of the proposed magnetic covert channel, a thread (or group of threads) that manipulate the CPU workload is tagged as suspicious. However, many kinds of applications use working threads that affect the processor workload, and this detection approach will likely suffer from a high rate of false alarms. Another problem is that the signal generation code involves only simple, non-privileged CPU operations (e.g., busy loops), without requiring special instructions or specialized API calls. Tracing non-privileged CPU operations at runtime requires that the processes be in step by step mode, which can severely degrade system performance [9].

Software-based detection also suffers from an inherent weakness in that it can be easily bypassed by malware. Static and dynamic detection of a malicious app in a smartphone is known to be a challenging task due to the wide range of code obfuscation and sandbox evasion techniques [43][44]. Moreover, apps that request access to the magnetic sensor are very common (e.g., for positioning and orientation), and therefore they cannot automatically be classified as malicious.

Another approach for detecting the covert channel is based on monitoring the magnetic field in the area of the computer. The magnetic field is analyzed to find hidden transmissions or deviations. Note that finding anomalies in the magnetic and electromagnetic spectrum may also suffer from a high rate of false positives [45][46].

B. *Prevention*

There are three different approaches that can be used to prevent attackers from establishing the proposed magnetic covert channel: shielding, jamming, and zoning.

**Shielding.** Magnetic shielding for computers is considered impractical except when used for special military or scientific purposes [47]. For effective magnetic shielding, special ferromagnetic materials such as mu-metal should be used [48]. However, even with ferromagnetic material it is difficult to provide effective magnetic shielding against low frequency magnetic fields [47].

**Signal Jamming**. In this approach, a strong signal is generated in the vicinity of the computer so that it interferes with unauthorized communications. Commercial magnetic field generators such as the compact magnetic field generator and analyzer MGA 1030 can generate a magnetic field at a strength of up to 1000 A/m at low frequencies (below to 1kHz) [49]. The power of such a magnetic field is hundreds of times stronger than the magnetic field generated by the CPU, so it overrides the CPU's magnetic signals. Software-level signal jamming involves the execution of background processes that initiate random magnetic transmissions, and these random signals will interfere with the transmissions of the malicious process. However, random workloads reduce system performance and may be infeasible in some environments.

**Zoning.** Procedural countermeasures involve the physical separation of emanating equipment from potential receivers. For example, the NATO SDIP-27 and SDIP-28 standards define separate zones in which electronic equipment is allowed [50]. In these standards, sensitive computers are kept in restricted areas in which certain equipment is banned. In our case, smartphones should be banned from close proximity to the sensitive computers.

IX. CONCLUSION

In this paper we show how attackers can exfiltrate information from air-gapped computers to nearby smartphones via magnetic signals. Recent work shows how magnetic signals can be generated from computers by regulating the CPU workload [24]. We introduce smartphones to the attack model and show that the magnetic signals can be received by a nearby smartphone via its magnetic sensors. We implement a malware that can control the magnetic fields emanating from the computer and evaluate the signal reception at various distances and bitrates. Our results show that the covert channel is effective for short distances (12.5cm) and operates at low bitrates (<5 bit/sec). However, the covert channel is effective in constrained environments where all wireless communication is blocked, e.g., when the smartphone is kept in a Faraday bag or configured in airplane mode. We also show that the malicious code can operate from within an isolated virtual machine and does not require special privileges for its execution.

X. REFERENCES

[1] M. Maybury, P. Chase, B. Cheikes, D. Brackney, S. Matzner, T. Hetherington, B. Wood, C. Sibley, J. Marin, and T. Longstaff, "Analysis and detection of malicious insiders," MITRE CORP BEDFORD MA, 2005.

[2] "Trump, Putin, and the New Cold War - The New Yorker." .

[3] S. Abraham and I. Chengalur-Smith, "An overview of social engineering malware: Trends, tactics, and implications," *Technology in Society*, vol. 32, no. 3, pp. 183–196, 2010.

[4] "Wikileaks: CIA uses 'Brutal Kangaroo' toolkit to hack air-gapped networks." .

[5] M. Guri, G. Kedma, A. Kachlon, and Y. Elovici, "AirHopper: Bridging the air-gap between isolated networks and mobile phones using radio frequencies," in *Malicious and Unwanted Software: The Americas (MALWARE), 2014 9th International Conference on*, 2014, pp. 58–67.

[6] M. G. Kuhn and R. J. Anderson, "Soft Tempest: Hidden Data Transmission Using Electromagnetic Emanations.," in *Information hiding*, 1998, vol. 1525, pp. 124–142.

[7] M. G. Kuhn, "Compromising emanations: eavesdropping risks of computer displays," University of Cambridge, 2002.

[8] M. Vuagnoux and S. Pasini, "Compromising Electromagnetic Emanations of Wired and Wireless Keyboards.," in *USENIX security symposium*, 2009, pp. 1–16.

[9] M. Guri, A. Kachlon, O. Hasson, G. Kedma, Y. Mirsky, and Y. Elovici, "GSMem: Data Exfiltration from Air-Gapped Computers over GSM Frequencies.," in *USENIX Security Symposium*, 2015, pp. 849–864.

[10] M. Guri, Y. Solewicz, A. Daidakulov, and Y. Elovici, "Fansmitter: Acoustic Data Exfiltration from (Speakerless) Air-Gapped Computers," *arXiv preprint arXiv:1606.05915*, 2016.

[11] M. Guri, Y. Solewicz, A. Daidakulov, and Y. Elovici, "Acoustic Data Exfiltration from Speakerless Air-Gapped Computers via Covert Hard-Drive Noise ('DiskFiltration')," in *European Symposium on Research in Computer Security*, 2017, pp. 98–115.

[12] M. Hanspach and M. Goetz, "On covert acoustical mesh networks in air," *arXiv preprint arXiv:1406.1213*, 2014.

[13] M. Guri, B. Zadov, and Y. Elovici, "LED-it-GO: Leaking (A Lot of) Data from Air-Gapped Computers via the (Small) Hard Drive LED," in *Detection of Intrusions and Malware, and Vulnerability Assessment: 14th International Conference, DIMVA 2017, Bonn, Germany, July 6-7, 2017, Proceedings*, M. Polychronakis and M. Meier, Eds. Cham: Springer International Publishing, 2017, pp. 161–184.

[14] M. Guri, D. Bykhovsky, and Y. Elovici, "aIR-Jumper: Covert Air-Gap Exfiltration/Infiltration via Security Cameras & Infrared (IR)," *arXiv preprint arXiv:1709.05742*, 2017.

[15] M. Guri, M. Monitz, Y. Mirski, and Y. Elovici, "Bitwhisper: Covert signaling channel between air-gapped computers using thermal manipulations," in *Computer Security Foundations Symposium (CSF), 2015 IEEE 28th*, 2015, pp. 276–289.

[16] "Tempest for Eliza." .

[17] M. Guri, M. Monitz, and Y. Elovici, "Bridging the Air Gap between Isolated Networks and Mobile Phones in a Practical Cyber-Attack," *ACM Transactions on Intelligent Systems and Technology (TIST)*, vol. 8, no. 4, p. 50, 2017.

[18] M. Guri, M. Monitz, and Y. Elovici, "USBee: Air-gap covert-channel via electromagnetic emission from USB," in *Privacy, Security and Trust (PST), 2016 14th Annual Conference on*, 2016, pp. 264–268.

[19] J. Loughry and D. A. Umphress, "Information leakage from optical emanations," *ACM Transactions on Information and System Security (TISSEC)*, vol. 5, no. 3, pp. 262–289, 2002.

[20] M. Guri, B. Zadov, A. Daidakulov, and Y. Elovici, "xLED: Covert Data Exfiltration from Air-Gapped Networks via Router LEDs," *arXiv preprint arXiv:1706.01140*, 2017.

[21] M. Guri, O. Hasson, G. Kedma, and Y. Elovici, "An optical covert-channel to leak data through an air-gap," in *Privacy, Security and Trust (PST), 2016 14th Annual Conference on*, 2016, pp. 642–649.

[22] A. C. Lopes and D. F. Aranha, "Platform-agnostic Low-intrusion Optical Data Exfiltration.," in *ICISSP*, 2017, pp. 474–480.

[23] N. Matyunin, J. Szefer, S. Biedermann, and S. Katzenbeisser, "Covert channels using mobile device's magnetic field sensors," in *Design Automation Conference (ASP-DAC), 2016 21st Asia and South Pacific*, 2016, pp. 525–532.

[24] Y. E. Boris Zadov Andrey Daidakulov Mordechai Guri, "ODINI : Escaping Sensitive Data from Faraday-Caged, Air-Gapped Computers via Magnetic Fields," 2018.

[25] V. P. Kodali, "Engineering electromagnetic compatibility: principles measurements technologies and computer models," 2001.

[26] M. N. Sadiku, *Elements of electromagnetics*. Oxford university press, 2014.

[27] R. Bansal, "Near-field magnetic communication," *IEEE Antennas and Propagation Magazine*, vol. 46, no. 2, pp. 114–115, 2004.

[28] Y. Cai, Y. Zhao, X. Ding, and J. Fennelly, "Magnetometer basics for mobile phone applications," *Electron. Prod.(Garden City, New York)*, vol. 54, no. 2, 2012.

[29] J. Lenz and S. Edelstein, "Magnetic sensors and their applications," *IEEE Sensors journal*, vol. 6, no. 3, pp. 631–649, 2006.

[30] "Industrial Defence In-Depth, Kaspersky Lab." .

[31] "The Epic Turla (snake/Uroburos) attacks | Virus Definition | Kaspersky Lab." .

[32] K. ZAO, "Red october diplomatic cyber attacks investigation." .

[33] "A Fanny Equation: 'I am your father, Stuxnet' - Securelist.".

[34] J. von Kistowski, H. Block, J. Beckett, C. Spradling, K.-D. Lange, and S. Kounev, "Variations in cpu power consumption," in *Proceedings of the 7th ACM/SPEC on International Conference on Performance Engineering*, 2016, pp. 147–158.

[35] "Sensor | Android Developers." .

[36] "download.yamaha.com/api/asset/file/?language=ja&site=jp.yamaha.com&asset_id=52799." .

[37] "Home | LibreOffice - Free Office Suite - Fun Project - Fantastic People." .

[38] "Official download of VLC media player, the best Open Source player - VideoLAN." .

[39] "Ubuntu Manpage: matho-primes - generate consecutive prime numbers." .

[40] "Some Thoughts on Faraday Bags and Operational Security | Micah Lee's Blog." .

[41] "Through-The-Earth Two-Way Emergency Wireless Communications for Mine Industry Safety." .

[42] "Core Motion | Apple Developer Documentation." .

[43] T. Vidas and N. Christin, "Evading android runtime analysis via sandbox detection," in *Proceedings of the 9th ACM symposium on Information, computer and communications security*, 2014, pp. 447–458.

[44] M. Guri, Y. Poliak, B. Shapira, and Y. Elovici, "JoKER: Trusted detection of kernel rootkits in android devices via JTAG interface," in *Trustcom/BigDataSE/ISPA, 2015 IEEE*, 2015, vol. 1, pp. 65–73.

[45] B. Carrara, "Air-Gap Covert Channels," Université d'Ottawa/University of Ottawa, 2016.

[46] S. Z. Goher, B. Javed, and N. A. Saqib, "Covert channel detection: A survey based analysis," in *High capacity optical networks and enabling technologies (HONET), 2012 9th international conference on*, 2012, pp. 57–65.

[47] "EMC for Systems and Installations Part 4 - Filtering and Shielding." .

[48] H. J. ter Brake, H. Wieringa, and H. Rogalla, "Improvement of the performance of a mu-metal magnetically shielded room by means of active compensation (biomagnetic applications)," *Measurement Science and Technology*, vol. 2, no. 7, p. 596, 1991.

[49] "Schloeder-EMV: Magnetic Field Generator and Analyzer." .

[50] R. Anderson, "Emission security," *Security Engineering,*, pp. 523–546, 2008.